\documentclass[aps,pre,preprint,showpacs,floatfix]{revtex4}
\usepackage{graphics,amssymb,amsmath,wasysym}
\begin{document}
\title{Modular synchronization in complex networks}
\author{E.~Oh$^1$, K.~Rho$^1$, H.~Hong$^2$ and B.~Kahng$^1$\\}
\affiliation{$^1$ School of Physics and Center for Theoretical Physics, \\
Seoul National University, Seoul 151-747, Korea\\
$^2$ Department of Physics, Chonbuk National University, Jeonju,
 561-756, Korea}
\date{\today}
\begin{abstract}
We study the synchronization transition (ST) of a modified
Kuramoto model on two different types of modular complex networks.
It is found that the ST depends on the type of inter-modular
connections. For the network with decentralized (centralized)
inter-modular connections, the ST occurs at finite coupling
constant (behaves abnormally). Such distinct features are found in
the yeast protein interaction network and the Internet,
respectively. Moreover, by applying the finite-size scaling
analysis to an artificial network with decentralized inter-modular
connections, we obtain the exponent associated with the order
parameter of the ST to be $\beta \approx 1$ different from
$\beta_{\rm MF} \approx 1/2$ obtained from the scale-free network
with the same degree distribution but the absence of modular
structure, corresponding to the mean field value.
\end{abstract}
\maketitle
Recently, there have been considerable efforts to understand
complex systems in terms of graph, consisting of vertices and
edges~\cite{review}. An interesting feature emerging in such
complex networks is a power-law degree distribution, $P_d(k)\sim
k^{-\gamma}$, where degree $k$ is the number of edges incident
upon a given vertex. Such networks are called scale-free (SF)
networks~\cite{ba}. Many SF networks in real world such as the
yeast protein interaction networks (PINs), the Internet, and
social networks contain modular structures within
them~\cite{ravasz,rives,sneppen,community}.
While recent studies have focused on the identification or
classification of such modules~\cite{newman}, little attention has
been paid to how such modules are interconnected each
other~\cite{jalan}. In this Letter, we study the synchronization
transition (ST) emerging from such modular SF networks. It is
found that the feature of the ST depends on the types of the
inter-modular connections, different from that occurring on the SF
network without modules.

Synchronization phenomena can be found in diverse
fields~\cite{sync,winfree}, which have been studied through
coupled oscillator models~\cite{review_sync}. Here, we study the
dynamics of synchronization on SF networks through a modified
Kuramoto model~\cite{kuramoto},
\begin{equation}
\frac{d\phi_i}{dt}=\omega_i -\frac{K}{k_i}\sum_{j~\in~{{\rm nn~of}~i}}
\sin(\phi_i-\phi_j), \label{ke}
\end{equation}
where $\phi_i$ and $\omega_i$ are the phase and the intrinsic
frequency of vertex $i$, respectively. $\omega_i$ is chosen from the
Gaussian distribution with unit variance. The summation runs over
$j$, the nearest neighbors (nn) of vertex $i$. Note that in contrast
to other works~\cite{review_sync,kuramoto}, the coupling strength is
used as $K/k_i$, where $k_i$ is the degree of vertex $i$. Due to the
factor $1/k_i$, the dephasing effect at the vertex with large degree
is reduced~\cite{motter}, and the ST occurs at finite $K_c$ in SF
networks even for $2 < \gamma < 3$~\cite{dslee}. Note that when the
factor $1/k_i$ is absent, $K_c=0$ for SF networks with $2< \gamma
<3$~\cite{dslee,ichinomiya}. The order parameter ${\cal M}$ is
defined as ${\cal M} \equiv \lim_{t\to \infty} \left\langle \Biggl|
\frac{1}{N}\sum_{j=1}^{N} e^{i\phi_j} \Biggr| \right\rangle$, where
$\langle\cdots\rangle$ is the ensemble average over different
configurations and $N$ is the total number of vertices. ${\cal M}$
is 0 (1) in the fully incoherent (coherent) phase.

\begin{figure}
\centering{\resizebox*{!}{4.5cm}{\includegraphics{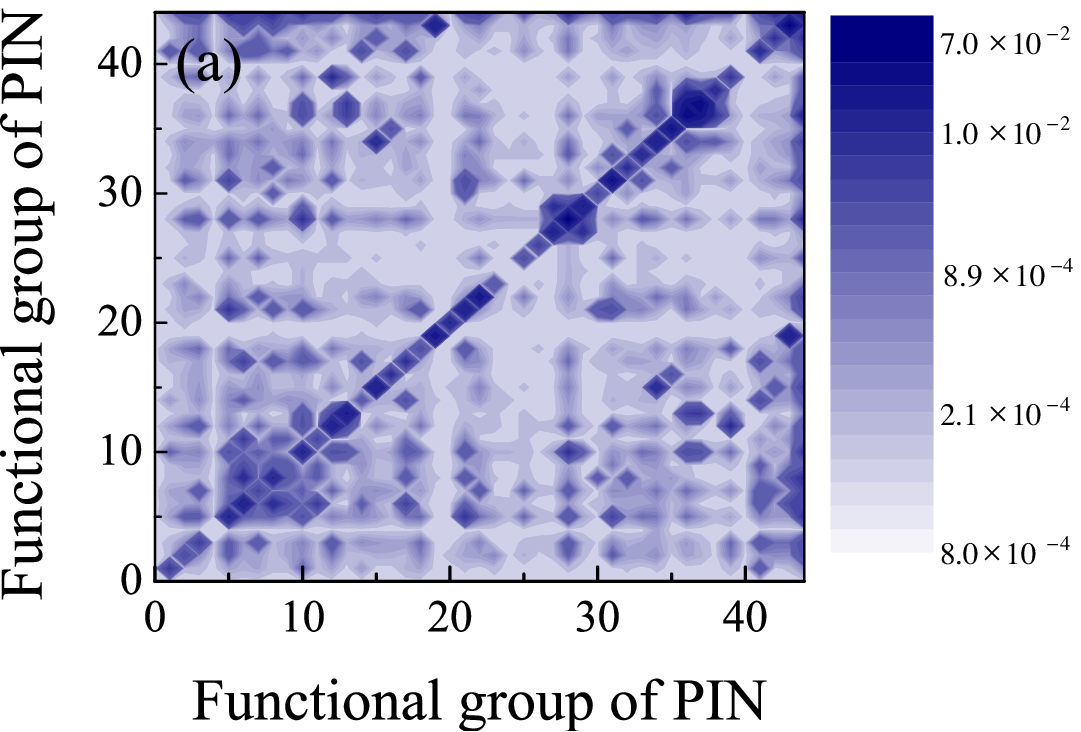}}}
\centering{\resizebox*{!}{4.5cm}{\includegraphics{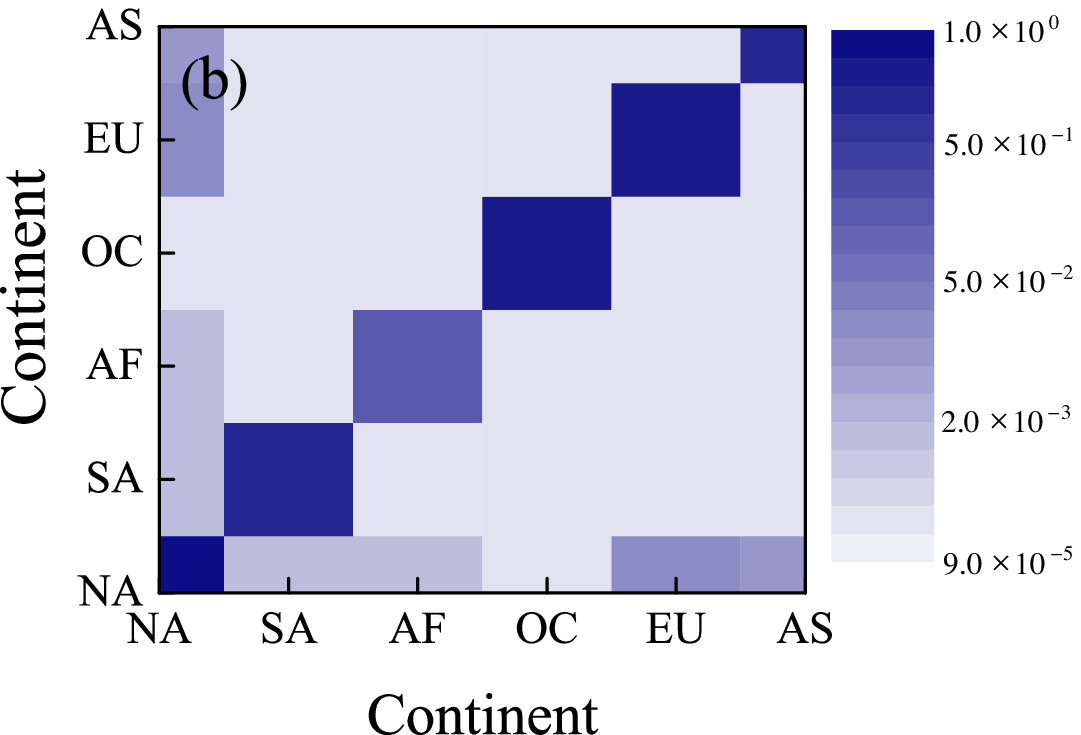}}}
\caption {(Color online). Profile of inter-modular connections 
for the PIN (a) and the Internet (b). The concentration
$c_{i,j}={\ell_{i,j}\ell_{j,i}/L_i L_j}$, where $\ell_{i,j}$ is
the number of edges emanating from module $i$ to module $j$ and the
$L_i=\sum_j \ell_{i,j}$.
NA, SA, AF, OC, EU, and AS stand for the North America, South America,
Africa, Oceania, Europe, and Asia, respectively.}
\label{fig:as_pin_group}
\end{figure}

\begin{figure}
\centering{\resizebox*{!}{4.5cm}{\includegraphics{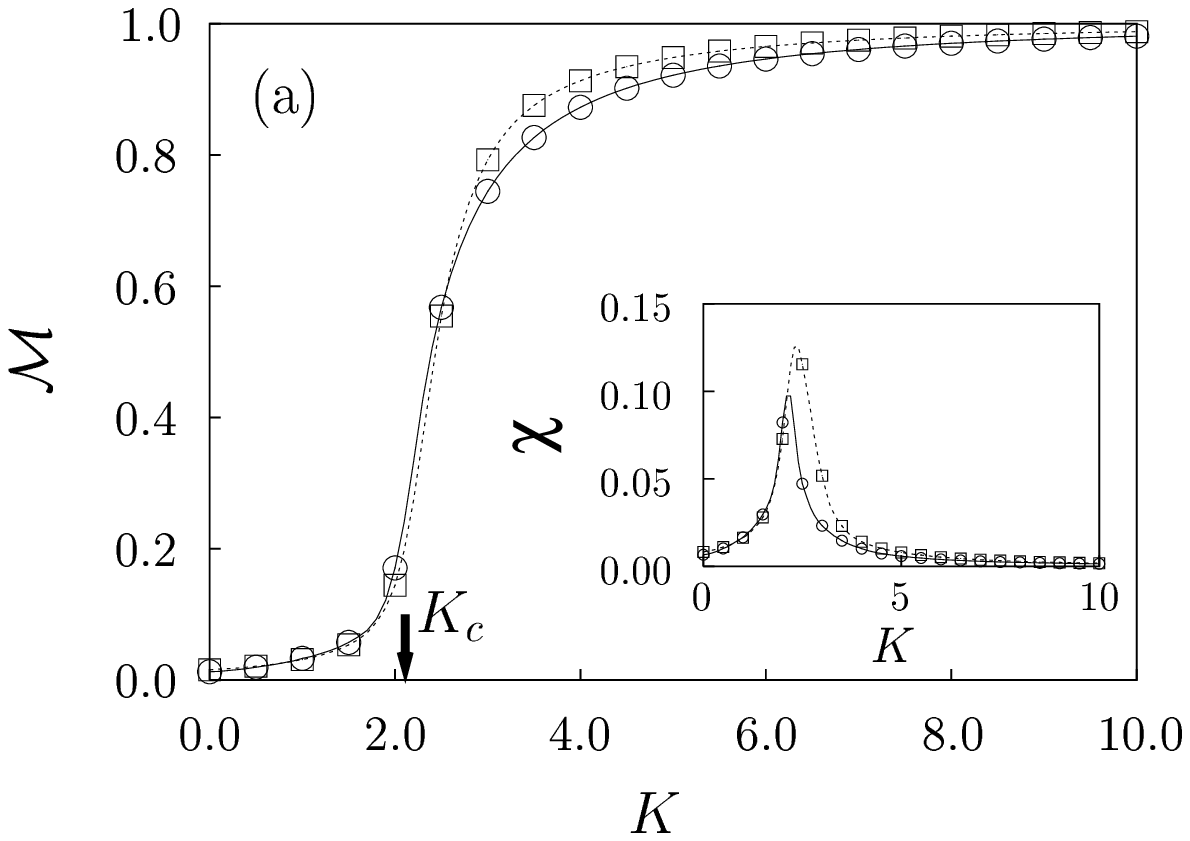}}}
\centering{\resizebox*{!}{4.5cm}{\includegraphics{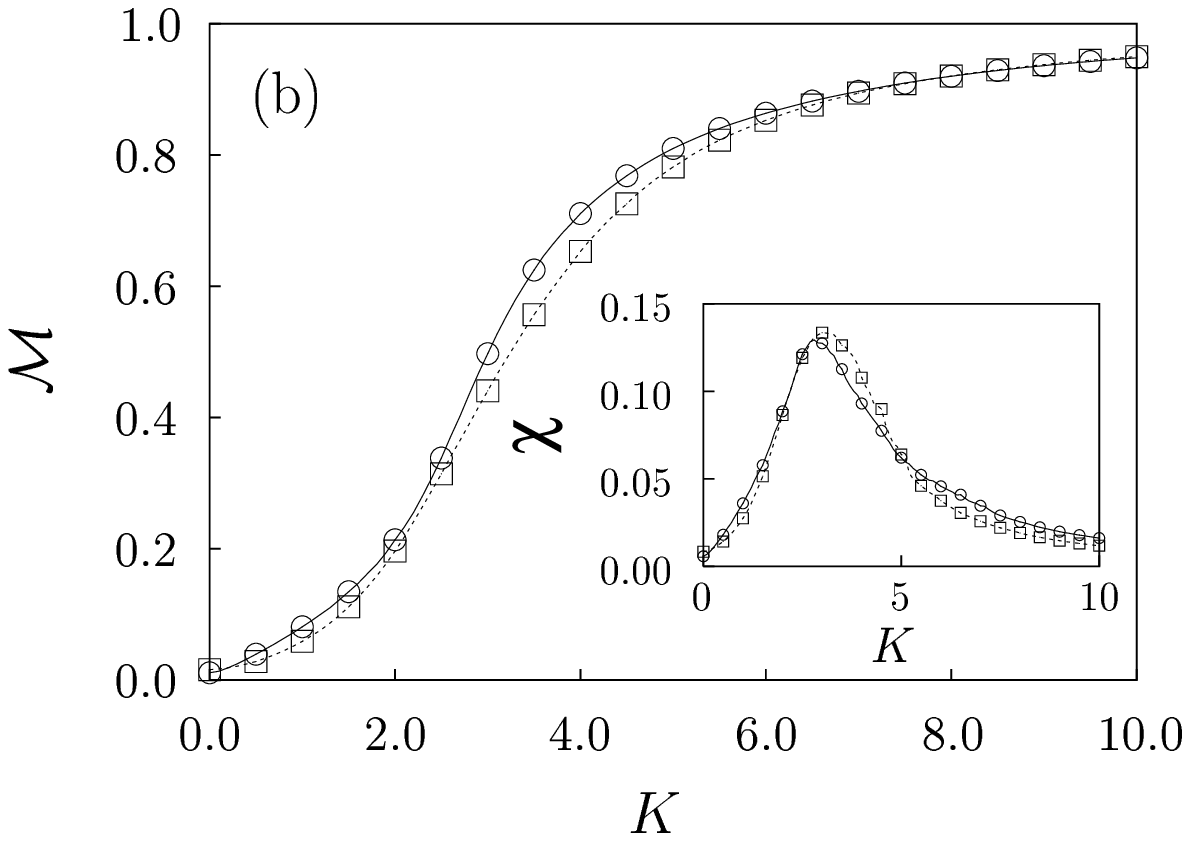}}}
\caption {Plot of the order parameter $\cal M$ as a function of
the coupling constant $K$ for two different types of complex
networks, the yeast PIN($\bigcirc$) and the CM ($\Box$) in (a)
and the Internet($\bigcirc$) and the hierarchy model
($\Box$) in (b). Insets : Plot of the susceptibility as a
function of $K$ for each case.}
\label{fig:as_pin_tr}
\end{figure}

To study how the profile of inter-modular connections affects the
ST pattern, we first investigate the inter-modular connections of
two real world networks, the yeast PIN and the Internet at
autonomous system level. Although the two networks are not
oscillator networks, we use them because they are prototypical
examples of modular networks and contain different types of
inter-modular connections. While a neuronal network is a good
example of oscillator networks, its structure has been still
veiled. As shown in Fig.~\ref{fig:as_pin_group}, the modules of
the PIN are interwoven diversely, while those of the Internet are
connected mainly through only a single module, corresponding to
the North America continent. The synchronization transition
patterns occurring on those two networks appear distinct in the
three aspects. First, as the coupling constant $K$ increases, the
order parameter increases very drastically for the former, while
it does gradually for the latter (Fig.~\ref{fig:as_pin_tr}).
Quantitatively, the order parameter increases from ${\cal M}=0.2$
to 0.8 by the increment $\delta K=1.3$ and $\delta K=2.9$ for the
PIN and Internet, respectively. Second, the susceptibility $\chi$,
defined as $\chi =\sqrt{\langle (r-\langle{r}\rangle)^2 \rangle}$
with $r=\lim_{t\rightarrow \infty}|({1/N})\sum_{j=1}^N
e^{i\phi_j}|$, exhibits a peak at the transition point $K_c$,
which is narrow (broad) for the former (latter) (the inset of
Fig.~\ref{fig:as_pin_tr}). To quantify it, we measure the
interquartile range~\cite{iqr}, which corresponds to the standard
deviation often used for the asymmetric distribution case, to be
1.3 and 3.0 for the PIN and Internet, respectively. Each number
means the interval of $K$ where 50\% of the data around the
peak position belong to. Third, when $K \gg K_c$, for the former, most
modules are synchronized with almost the same phase, while for the
latter, individual modules are synchronized independently, leading
the overall system to be weakly synchronized. We also confirm such
different types of the synchronization patterns through two
artificial networks below.

Two artificial networks are defined as follows: One is a
modification of the community model (CM), proposed by Girvin and
Newman~\cite{newman}. There, $n_c$ modules are present from the
beginning, which is fixed under evolution, while the total numbers
of vertices and edges grow with time. At initial, each module
contains $m$ vertices and they are fully connected. At each time
step, a new vertex is added to the system and chooses a module at
random. The newly added vertex attaches $m_0$ edges to existing
vertices selected with the probability proportional to its
degree~\cite{ba}, belonging to the chosen module. This process is
repeated until the step when each module contains $n_0$ vertices
on average. Then $n_c$ sets of distinct SF networks are generated.
After those steps, we add $m_e$ edges, connecting vertices
belonging to different modules, which are selected randomly. The
parameters, $n_c$, $n_0$, $m_0$ and $m_e$ are adjustable. To
compare the resulting network with the yeast PIN, we take $n_c=5$
and $n_0=625$, $m_0=3$ and $m_e=934$, making the mean degree be
6.58, comparable to that of the PIN, 6.55. While the CM may be too
artificial or have too many parameters, it simply reflects the
presence of modules and their diverse connections. The second one
is the hierarchy model introduced by Ravasz and
Barab\'asi~\cite{hierarchy}. The model contains 5 modules at the
highest level, each of which contains 5 submodules within it in a
hierarchical manner. The inter-modular interactions in the
hierarchy model are made only through one particular module among
the 5 modules at the same hierarchical level. The mean degree of
the hierarchy model is $\langle k \rangle=7.8$.

\begin{figure}
\centering{\resizebox*{!}{5.0cm}{\includegraphics{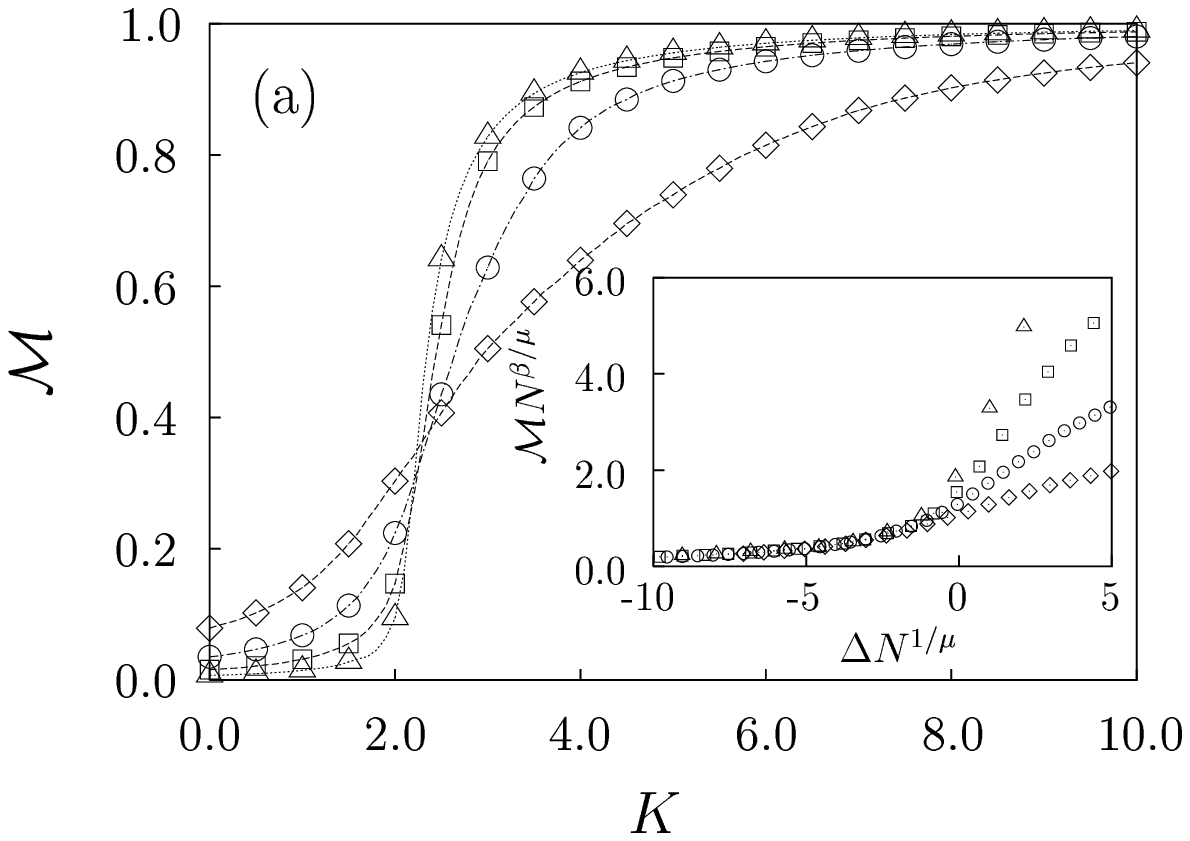}}}
\centering{\resizebox*{!}{5.0cm}{\includegraphics{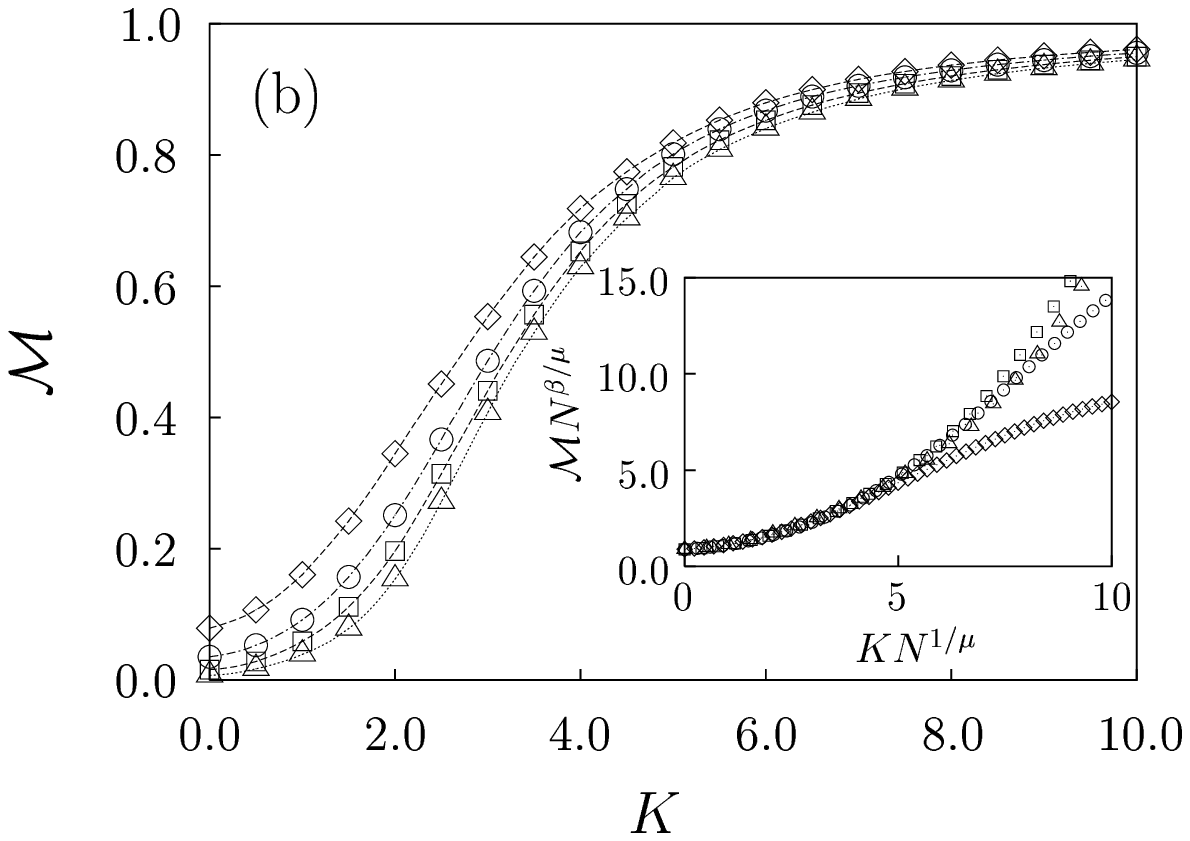}}}
\caption {Size dependence of ${\cal M}$ in the CM (a) and the hierarchy
model (b). The data are obtained for different system sizes
125($\Diamond$), 625($\bigcirc$), 3125($\Box$) and 15625($\triangle$)
with 1000 configurations for each. Inset: Plot of the order parameter
in the scaling form.}
\label{fig:size_gn_rb_tr}
\end{figure}

The distinct ST patterns can be observed more obviously in the
size-dependent behaviors of ${\cal M}$ in the CM and the hierarchy
model. The system size $N$ is controlled by changing the number of
vertices in each module $n_0$ in the CM and the number of
hierarchical levels in the hierarchy model. As shown in
Fig.~\ref{fig:size_gn_rb_tr}, for the CM, the ST occurs at $K_c
\simeq 2.1$, which is the crossing point of ${\cal M}$ for
different sizes. We find numerically that the order parameter
scales as ${\cal M} \sim N^{-\beta/\mu}F(\Delta N^{1/\mu})$, where
$\Delta=K-K_c$, $\beta/\mu \simeq 0.25$ and $1/\mu \simeq 0.25$.
Thus $\beta \approx 1$ for the CM network. To check if it is
related to the existence of the modular structure, we destroy the
modular structure in the CM model by rewiring edges. Two different
edges are randomly selected and their one-end vertices are
exchanged unless this swap procedure makes the network
disconnected. The swapping process is applied as many as two times
of the total number of edges. Then the resulting network loses
modules but the degree distribution remains the same. Applying the
finite-size scaling analysis to the swapped CM network, we obtain
the mean field value $\beta_{\rm MF} \approx 1/2$. Thus the nature
of the ST is changed by the absence of modular structure. On the
other hand, for the hierarchy model, we cannot find any crossing
point in ${\cal M}$ for different sizes. For small $K$, the data
are well fit to the scaling form of ${\cal M}\sim
N^{-\beta/\mu}F(KN^{1/\mu})$ with $\beta/\mu \approx 0.5$ and
$1/\mu \simeq 0.17$ (Fig.~\ref{fig:size_gn_rb_tr}b).

\begin{figure}
\centering{\resizebox*{!}{5.0cm}{\includegraphics{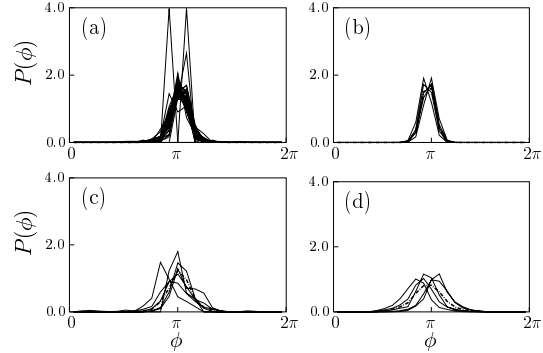}}}
\caption {The normalized phase distribution of each module (solid line)
at the coupling
constant $K=7.0$ for the PIN (a), the CM (b), the Internet (c),
and the hierarchy model (d). The dashed line represents the normalized
phase distribution of the overall system.}
\label{fig:phase_dis}
\end{figure}

\begin{figure}
\centering{\resizebox*{!}{5.0cm}{\includegraphics{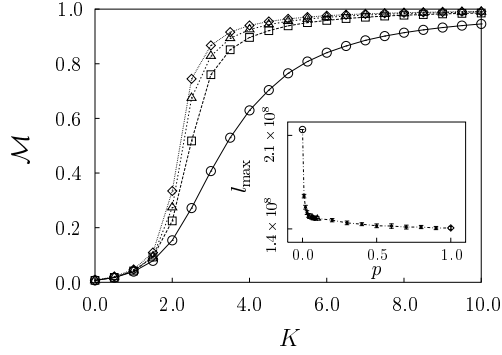}}}
\caption {The interplay of the ST pattern in the hierarchy model
as a function of $p$, the swapping ratio, for $p=0$ ({\Large
$\circ$}), $p=0.1$ $(\square)$, $p=0.5$ $(\triangle)$, and $p=1.0$
$(\diamond)$. Inset: Plot of $\ell_{\rm max}$ as a function of
$p$.} \label{fig:trans}
\end{figure}

To understand the emergence of the different ST patterns from a
microscopic respect, once we identify modules of each network and
examine the synchronization within each module. For the yeast PIN,
43 modules are identified according to their functional categories
and for the Internet, six modules are done according to the
continents in the world. Modules for the artificial networks can
be naturally identified. In Fig.~\ref{fig:phase_dis}, we plot the
phase distributions at $K=7.0$ for the four networks. Each curve
represents the normalized phase distribution of each module. For
the PIN, the phase distributions of each module overlap each
other, but the three modules, cell adhesion, mitochondrial
transcription and septation. For the CM, most of the phase
distributions overlap with each other, implying that the networks
are globally synchronized. For the Internet and the hierarchy
model, however, the phase distribution functions of each module do
not overlap, indicating that each module is synchronized
independently. The network is modularly synchronized. Thus, the
global synchronization for the Internet and the hierarchy model
appears weaker than that for the PIN and the CM for the given $K$.

To see the interplay between the two ST patterns, we deform the
network structure of the hierarchy model by swapping edges as
used in the CM network. The ratio $p$ between the
number of swapping edges and the total number of edges is a
control parameter. With increasing $p$, the inter-modular
connections are decentralized. As can be seen in
Fig.~\ref{fig:trans}, the ST pattern becomes sharper as the ratio
$p$ increases, and in particular it changes drastically up to $p
\approx 0.1$, beyond which they look very similar and overlap
with the ST data for the PIN or the CM.

\begin{figure}
\centering{\resizebox*{!}{5.0cm}{\includegraphics{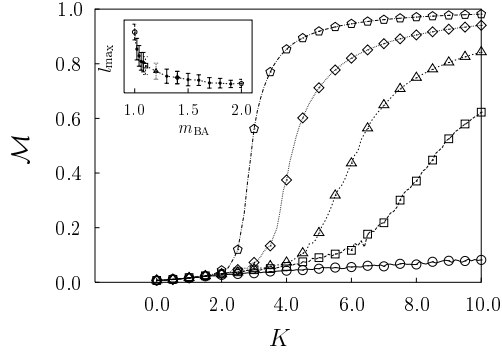}}}
\caption {The ST patterns as a function of $K$ for various
mean numbers of emanating edges of the BA model, $m_{\rm BA}= 1.0$
$({\Large \circ})$, $1.1$ $(\square)$, $1.2$ $(\triangle)$, $1.4$
$(\diamond)$, and $2.0$ $(\pentagon)$.
Inset: Plot of $\ell_{\rm max}$ as a function of $m_{\rm BA}$.}
\label{fig:chim}
\end{figure}

It was recently shown that the synchronizability in complex
networks is related to the maximum value of load
$\ell_{\rm max}$~\cite{lai,hong04}. Load is the effective amount
of data packets passing through a given vertex when every pair of
vertices sends and receives a unit data packet along the shortest
pathways between them~\cite{load}. To examine how $\ell_{\rm max}$
is related to the synchronization pattern, we investigate
$\ell_{\rm max}$ as a function of the swapping ratio $p$. As shown
in the inset of Fig.~\ref{fig:trans}, $\ell_{\rm max}$ decreases
rapidly in the range of $0 < p < 0.1$, however, it decreases very
slowly or is almost constant for $p > 0.1$. This suggests that
across $p\approx 0.1$, the synchronizability becomes enhanced
drastically, in accordance with the behavior of the ${\cal M}$.

It would be interesting to see such a crossover through the
Barab\'asi and Albert (BA) model. We control the number of edges
$m_{\rm BA}$ emanating from a newly added node in the BA model.
When $m_{\rm BA}=1$, ${\cal M}$ is almost flat with changing $K$.
For larger $m_{\rm BA}$, ${\cal M}$ changes more sharply with
increasing $K$ as shown in Fig.~\ref{fig:chim}.
$\ell_{\rm max}$ decreases
drastically up to $m_{\rm BA} \approx 1.1$ and gradually decrease
beyond that as shown in the inset of Fig.~\ref{fig:chim},
reflecting that the synchronizability enhances drastically at
$m_{\rm BA}\approx 1.1$. This fact is related to the topological
feature as follows. When $m_{\rm BA}=1$, the network is tree.
As $m_{\rm BA}$ increases, the edges connecting different branches
are formed, and its fraction is non-negligible for $m_{\rm BA}\ge 1.1$.
As a result, the synchronization transition occurs more drastically
for $m_{\rm BA}\ge 1.1$.
It was found that as $m_{\rm BA}$ increases, the exponent $\delta$
associated with the load distribution, $P_{\ell}(\ell)\sim \ell^{-\delta}$,
increases drastically from $\delta \approx 2.0$, and almost
remains as 2.2 for $m_{\rm BA} \ge 1.1$~\cite{ghim}.
This fact is closely related to the crossover behavior of ${\cal M}$.

In summary, we have studied the two different types of the
synchronization patterns on SF networks through a modified Kuramoto
model. Such different patterns are mainly rooted from different
types of inter-modular connection profile. When the inter-modular
connections are decentralized (centralized), the ST occurs very
drastically (gradually) and the synchronization arises globally
(within modules), of which the behavior was analyzed quantitatively.
We also showed that the ST point occurs at finite coupling constant
for the former, while it is not well identified for the latter by
performing finite-size scaling analysis. The critical exponent
$\beta$ associated with the order parameter is obtained to be
$\beta\approx 1$ for the former case. The crossover behavior between
the two cases was also studied by rewiring edges from the hierarchy
model and by modifying the BA model.
Finally, it is interesting to note that the global or the modular
synchronization may be related to epileptic seizure. While the
brain is electrically activated, any disruption within it may
cause abnormal functioning. Then neurons in one hemisphere misfire
and generate abnormal electrical activities, which spread to the
other hemisphere through diverse pathways, leading the global
synchronization~\cite{conlon}. To treat it, the corpus callosum,
the main communication pathways between the two hemispheres, is
sometimes cut to prevent the communications from one side to the
other. It reveals through recent experiment~\cite{vergnes} that
while the synchronization after the cut arises within one module
mostly, the global synchronization also arises occasionally, which
is accomplished via other pathways.

This work is supported by the KOSEF Grant No. R14-2002-059-01000-0
in the ABRL program.

\end{document}